# Monte Carlo study of coaxially gated CNTFETs: capacitive effects and dynamic performance

H. Cazin d'Honincthun[1,2], S. Galdin-Retailleau[1], A. Bournel[1], P. Dollfus[1], J.P. Bourgoin[2]

[1]IEF, **CNRS, Univ. Paris-Sud, Bât. 220, 91405, Orsay, France**
[2]**Molecular Electronics Laboratory**, SPEC, CEA Saclay, 91191, Gif-sur-Yvette, France

**Corresponding author**
Hugues Cazin d'Honincthun
Institution: Institut d'Electronique Fondamentale,
CNRS UMR 8622, Université Paris Sud, Bât. 220
F-91405 Orsay cedex France
E-Mail: hugues.cazin@ief.u-psud

## Abstract

Carbon Nanotube (CNT) appears as a promising candidate to shrink field-effect transistors (FET) to the nanometer scale. Extensive experimental works have been performed recently to develop the appropriate technology and to explore DC characteristics of carbon nanotube field effect transistor (CNTFET). In this work, we present results of Monte Carlo simulation of a coaxially gated CNTFET including electron-phonon scattering. Our purpose is to present the intrinsic transport properties of such material through the evaluation of electron mean-free-path. To highlight the potential of high performance level of CNTFET, we then perform a study of DC characteristics and of the impact of capacitive effects. Finally, we compare the performance of CNTFET with that of Si nanowire MOSFET.

## Résumé

**Influence des effets capacitifs sur les performances dynamiques d'un CNTFET par la méthode Monte Carlo.** Le nanotube de carbone (CNT) est à ce jour l'un des candidats les plus prometteurs pour faire passer le transistor à effet de champ (FET) à l'échelle du nanomètre. Des recherches intensives sont en cours afin de déterminer les caractéristiques statiques et dynamiques des transistors à nanotube de carbone (CNTFET). Nous présentons dans cette étude des résultats de simulations de CTNFET par la méthode Monte-Carlo avec prise en compte des interactions électrons-phonons. Un des objectifs est de présenter les propriétés du transport pouvant être atteintes dans ce matériau par une évaluation du libre parcours moyen des porteurs. Une étude des caractéristiques statiques du CNTFET est réalisée et permet de mettre en avant l'influence du contrôle capacitif par la grille sur les performances. Enfin nous comparons les performances d'un CNTFET avec celles obtenues par simulation d'un transistor à nanofil de Silicium.

## 1. Introduction

Field-effect transistors (FETs) based on semiconducting carbon nanotubes (CNTs) have generated considerable interest in the past few years because of their quasi ideal electronic properties and have reached recently a high level of performance [1-7]. Today, both p- and n-type CNT field-effect transistors (CNFETs) have been fabricated and have exhibited promising characteristics. However, the device physics and transport mechanisms are not yet fully understood and important efforts have been put on modelling CNTFETs to optimize their ultimate device performance capability.

Among numerous remarkable properties of CNTs one can mention their 1D nature which yields a very good charge confinement. The electrostatic control is thus all the better that CNTs may be conveniently processed with high-k gate oxide. Moreover, in contrast to silicon, transport properties in this material are excellent and nearly similar for holes and electrons. In semiconducting nanotubes, mean free paths of 300-500 nm at low field and 10-100 nm at high field have been measured [8-9] and many works put forward their high potential of ballistic transport. Finally, the combination of metallic and semiconducting CNTs makes possible to think of a full-nanotube circuit design [10-11].

In the present work we use our Monte Carlo code MONACO [12,13] to study (i) the intrinsic transport properties of semiconducting CNTs and (ii) the operation and dynamic behaviour for an all-around gated CNT with gate length of 100 nm and ohmic source/drain contacts. All relevant electron-phonon scattering mechanisms are taken into account. The study is focused on the potential of ballistic transport in CNTs according to the electric field and tube diameter and on the influence of the gate capacitance on static and dynamic characteristics for CNTFETS which may operate close to the quantum capacitance limit. We present a detailed comparison between CNTFET and Si MOSFET performances.

## 2. Description of nanotube electrical properties and transport model

The energy dispersion of the nanotube can be obtained using numerical tight-binding calculation including tube curvature effects and the influence of mixing of in-plane $\sigma$ and out of plane $\pi$ carbon orbitals. This effect becomes very important for low diameter tubes. The unit cell of a zigzag CNT ($n,0$) contains $4n$ atoms with 4 orbitals per atom, which leads to $16n$ valence and conduction subbands. Here we consider the three lowest subbands which are twofold degenerate in two valleys, each of them centered on a grapheme K point. In the energy range 0-1.5 eV, the energy dispersion $E_p(k_z)$ of subband $p$ is conveniently fitted by:

$$\frac{\hbar^2 k_z^2}{2 m_p} = \left[E_p - E_p^m\right]\left[1 + \alpha_p\left(E_p - E_p^m\right)\right] \qquad (1)$$

where $E_p^m$, $m_p$, and $\alpha_p$ are the minimum of energy, the effective mass and the non-parabolicity coefficient of the subband, respectively.

Phonons are the main source of electron scattering in carbon nanotube. The phonon spectrum of CNTs may be derived from that of graphene using the zone-folding method [14]. This method breaks each of the six phonon branches of graphene into $2n$ sub-branches. Longitudinal acoustic and optical modes are considered to be dominant for electrons scattering within the first three subbands [15]. For acoustic modes that generate intra-subband scattering, we use the same phonon energies as in [16]. For inter-valley and inter-subband acoustic phonon scattering, we consider the branches to be flat with a constant energy equal to the minimum value of each mode. Two phonon modes resulting from the longitudinal optical branch of graphene are considered with energy values of 180 meV and 200 meV [17]. They induce inter-valley and intra-valley scattering. Finally, we also include the radial breathing phonon mode which is typical of single-wall carbon nanotubes [18,19].

The electron-phonon scattering rates are calculated using the Fermi golden rule within the deformation potential model. Intra-subband acoustic scattering is treated as an elastic process while inter-subband transitions are considered through associated finite phonon energy [13].

## 3. Electron tansport

### 1. Steady-state results

Here we present some transport properties in perfect CNTs, with diameter $d_t$ ranging from 0.8 nm to 4.6 nm, i.e. for a wrapping index $n$ ranging from 10 to 59. Transport calculations have been performed at room temperature under uniform field applied along the tube axis. We first calculated some steady-state velocity-field characteristics (not shown). The velocity is found to increase with the tube diameter $d_t$, and the peak velocity reaches $3.43 \times 10^7$ cm/s for $n = 10$ ($d_t = 0.783$ nm) and $4.88 \times 10^7$ cm/s for $n = 58$ ($d_t = 4.62$ nm). Similar results have been reported [16,20]. These high values of electron saturation velocity have to be compared with the value of $9.6 \times 10^6$ cm/s in bulk silicon.

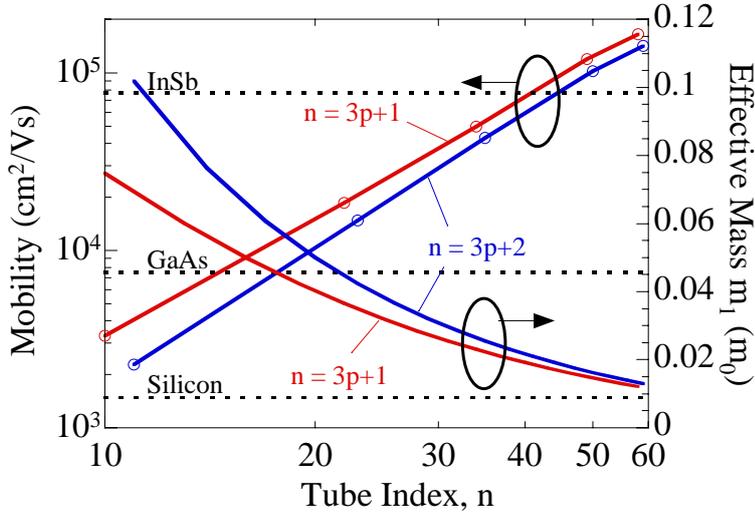

**Fig. 1. Effective mass $m_1$ in the first subband and electron mobility as a function of wrapping index $n$ for two types of semiconducting zigzag CNTs; Silicon and III-V material mobilities are reported for comparison (at 300 K).**

The low-field mobility is plotted in Fig. 1 as a function of the tube index. The results are separated in two different curves depending on whether $n+1$ or $n+2$ is multiple of 3, which is in accordance with the evolution of the effective mass $m_1$ in the first subband. The mobility increases from 2 300 cm²/Vs for small diameter to 142 000 cm²/Vs for large diameter, which is consistent with the reduction of effective mass $m_1$. These high values exceed those of any usual semiconductor, even InSb with the highest-known mobility at room temperature, and seem consistent with available experimental measurements which exhibit mobility values as high as 80 000 cm²/Vs [3].

In Fig. 2, we plot together the subband occupation and the electron velocity as a function of field for CNTs of index $n = 22$ ($d_t = 1.72$ nm, solid lines) and $n = 49$ ($d_t = 3.83$ nm, dashed lines). For the small diameter tube, a monotonous increase of velocity, as a function of field, is observed up to the saturation peak velocity. Otherwise, for the large diameter tube, the velocity-field curve is characterized by a change of slope and a shoulder, at low-field (about $E \approx 0.1$ kV/cm). These different behaviours can be explained by the difference in subband energy splitting which is reflected in the threshold energy of inter-subband scattering rates. It should be noticed that all changes of slope in the velocity-field curves in Fig. 2 are strongly related to electron transfer from the first to the second subband where the effective mass $m_2$ is higher than $m_1$. Consider first the large diameter CNT ($n = 49$), for which the energy distance between the two first subbands ($\Delta E_{12} = 115$ meV) is smaller than the smallest energy of intervalley phonons ($\hbar\omega = 160$ meV). The first change of slope (shoulder) in the velocity-field curve is due to transfer from first to second subband of the same valley by emission of low-energy phonon ($\hbar\omega = 7.5$ meV) which occurs at lower energy than intervalley transition. The resulting increase in effective mass influences the velocity, but the transfer rate is too small for this mechanism to induce the velocity saturation. Actually, the velocity peak and saturation only occur when intervalley intersubband transfer by high energy phonon emission becomes possible and further increases the electron population in the second subbands. Now consider the small diameter tube ($n = 22$), the energy splitting between the two first subbands is higher ($\Delta E_{12} = 265$ meV) than for the latter. So high rate intervalley emission ($\hbar\omega = 160$ meV) occurs at lower energy than intravalley intersubband transitions, whose effect on the velocity-field curve is now negligible. Again, the velocity peak and velocity saturation occur when inter-valley inter-subband transfer becomes possible. The critical field $E_{cr}$ associated with the peak velocity is slightly higher for the small diameter tube ($E_{cr} \approx 11$ kV/cm) than for the large one ($E_{cr} \approx 7$ kV/cm), because of higher $\Delta E_{12}$ splitting. For higher electric field, we observe a slight velocity decrease due to electron transfer to higher subbands with higher effective masses subbands. This phenomenon was also observed by other groups [18,20].

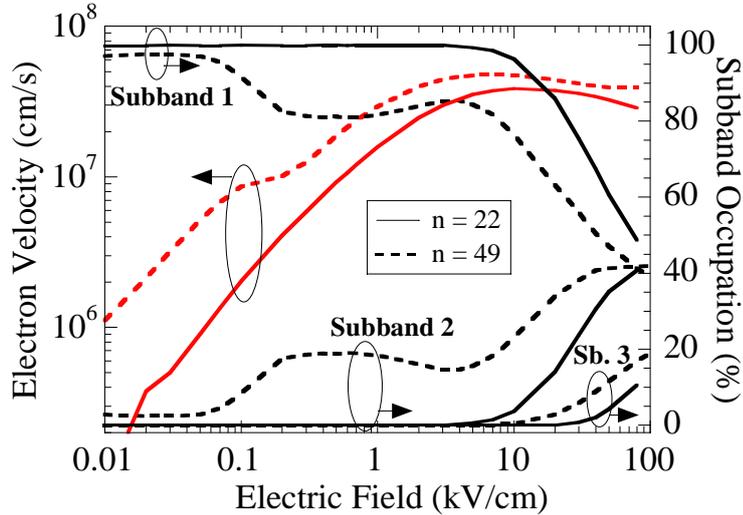

**Fig. 2 Steady-state velocity and subband occupation as a function of field for $n = 22$ and $n = 49$.**

### 2. Transient transport and ballisticity

Now, considering an electron gas initially at equilibrium under $E = 0$, we study its transient behaviour in response to a field step of height $E_s$ applied at time $t = 0$. We use the possibility of counting the number of scattering events experienced by each carrier between the positions $z = 0$ and $z = L$ [21]

We have access to the number of purely ballistic electrons as a function of the length $L$, as shown in Fig. 3 for various CNTs with a field step of 1 kV/cm. For this low field regime, the main scattering event corresponds to electron-phonon acoustic scattering. The curves decrease slowly for large tubes ($n \geq 49$, $d_t \geq 3.8$) with 50% of electrons which are still ballistic at $L = 1\,\mu m$. For small tubes, the electron ballisticity decreases rapidly on the first 200 nm. The probability to have ballistic electrons on a given distance is all the higher that the diameter is larger, which is strongly related to the diameter-dependence of effective mass.

The influence of electric field on ballistic transport is illustrated in Fig. 4 where we plot the fraction of ballistic electrons for $n = 22$. If we consider the length for which this fraction is equal to 50%, Fig. 4 shows that this length is about 250 nm for a field of 6 kV/cm and it is only 75 nm for a field of 30 kV/cm. It even falls to 60 nm under a field of 60 kV/cm (not shown). For each curve, there is an abrupt fall in the fraction of ballistic electrons. The length $L_F$ at which this fall occurs corresponds to the distance needed for a ballistic electron to reach the minimum kinetic energy which makes possible an inter-valley phonon emission ($\hbar\omega = 160$ meV and 180 meV). Thus $L_F$ is roughly defined by $k_B \times T/2 + E \times L_F \approx \hbar\omega$. As soon as this phonon emission is possible, the emission rate is so high that the ballistic probability vanishes rapidly. This phenomenon explains the apparent absence of abrupt fall for the field of 1 kV/cm, since the distance needed to reach 160 meV is around 1.5 µm. However, low energy carriers drifted in this low field, are thus kept under the influence of electron scattering by quasi-zero energy acoustic phonons which explains the strong electron ballisticity decrease in the first 100 nm.

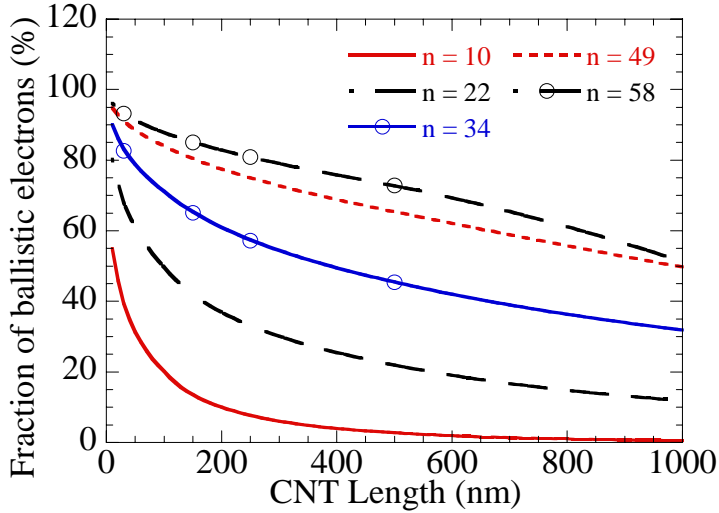

**Fig. 3. Fraction of ballistic electrons as a function of CNT length (E =1 kV/cm).**

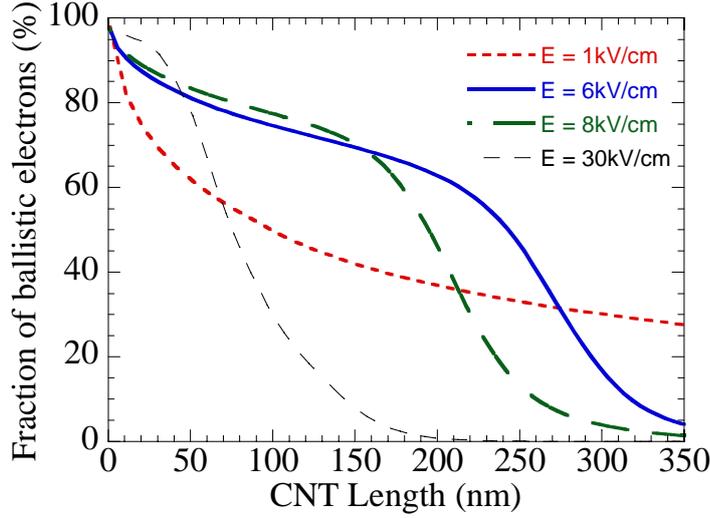

**Fig. 4. Fraction of ballistic electrons as a function of CNT length for various electric field ( *n* = 22 ).**

We also extracted the electron mean free path MFP associated with the different phonon scattering processes as a function of electric field. We plot in Fig. 5 the results obtained for three wrapping indices by considering separately the two main scattering mechanisms, i.e. elastic intra-subband scattering via low-energy acoustic phonons and inelastic intervalley scattering via high energy optical phonons. The MFP for elastic scattering is weakly dependent on field but ranges from 100 nm to 600 nm according to the tube diameter. In contrast, the MFP for intervalley scattering is almost independent on diameter but strongly decreases at high field as a consequence of carrier heating. It is only 20 nm for a field of 100 kV/cm, in agreement with other works [10,11]. These results clearly show the interest in using CNTs in the low-field regime [1-10 kV/cm] to fully benefit from a high potential of ballistic transport only limited by quasi-elastic scattering with low rate.

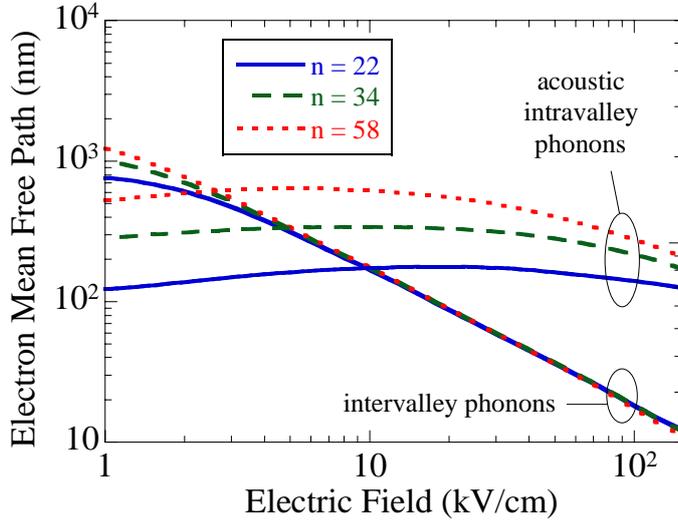

**Fig. 5. Electron mean free path due to acoustic and intervalley phonon scattering as a function of electric field for *n* = 22, 34, 58.**

## 4. Transistor

### 1. Device modelling

For device simulation, the Monte Carlo (MC) transport algorithm is self-consistently coupled with a 3D Poisson solver based on a finite-element scheme. In this work, Pauli blocking is taken into account by using the Lugli rejection method [22]. Before each scattering event, a random number is compared to the local carrier distribution function of the selected final state to decide whether the scattering is accepted or rejected and treated like a self-scattering event. With this technique, no scattering can occur if the final energy state is occupied.

In CNTFETs source and drain contacts are either Schottky or ohmic-like. In this work, ohmic boundary conditions are considered at source and drain contacts and are assumed to be characterized by thermal equilibrium and local charge neutrality. After each time step, the appropriate number of electrons is injected at each ohmic contact to recover this neutrality condition. The wave vector coordinates of injected carriers are randomly sampled from a Fermi-Dirac distribution.

Considering the CNT (19,0) of diameter $d_{tube} = 1.5$ nm and of energy gap 0.55 eV, we used our simulator to examine an all-around gated CNTFET with 20 nm-long n-doped extensions at source and drain ends. First, these extensions provide charge carriers, like in conventional MOSFETs. Then, they allow recovering the experimental situation where carrier density is controlled either electrostatically through multiple gates [23] or by charge transfer by depositing electron-donors adsorbed on them, like K atoms [24] or molecules [25]. Here, extensions are uniformly doped to $5 \times 10^6$ cm$^{-1}$ which corresponds to 0.0014 dopant per carbon atom.

The channel is 100 nm-long and is surrounded by an HfO$_2$ layer with dielectric constant $\varepsilon = 16$. We consider five equivalent SiO$_2$ oxide thicknesses (EOT) of 5.3, 1.4, 1, 0.4 and 0.2 nm. It corresponds to oxide capacitance $C_{OX}$ values of 112, 225, 280, 560 and 1120 pF/m, respectively. This large range of simulated capacitances allows covering almost the whole range of high-κ dielectric thicknesses which can be processed for such devices.

## 2. Static performances

Figures 6 and 7 show output characteristics $I_D$-$V_{DS}$ for a low EOT value (0.2 nm) for different gate-to-source voltages and $I_D$-$V_{GS}$ characteristics for the five $C_{OX}$ values and under a drain-to-source voltage $V_{DS} = 0.7$ V, respectively. The former exhibits high conductance values in ohmic regime, and low values in linear regime, performed by a strong channel electrostatic gate control. Fig. 7 shows an expected $C_{OX}$-induced current enhancement that is however limited at high $C_{OX}$ as explained below. In both figures, at high $V_{GS}$ and low EOT, we observe a strong drain current saturation. This behaviour can be explained with Fig. 8, which shows the 1st subband profile for two $C_{OX}$ values and different $V_{GS}$ at $V_{DS} = 0.4$ V. By considering a given gate voltage $V_{GS} = 0.4$ V, a source access resistance is present for $EOT = 0.2$ nm in contrast to $EOT = 5.3$ nm. It interestingly corresponds to the voltage regime where the drain current begins to saturate for the lower EOT in Fig. 7 which is clearly due to a source access resistance limitation.

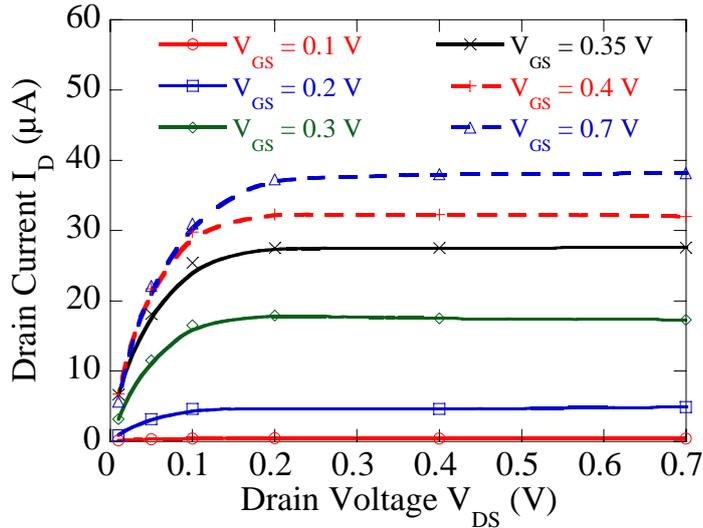
**Fig. 6.** Output characteristics $I_D$-$V_{DS}$ for EOT = 0.2 nm.

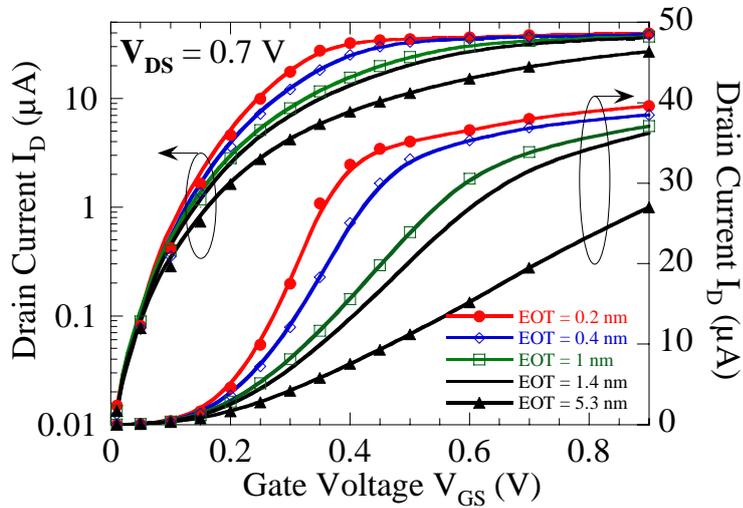
**Fig. 7.** $I_D$-$V_{GS}$ characteristics for 5 oxide capacitances and $V_{DS}$ = 0.7 V.

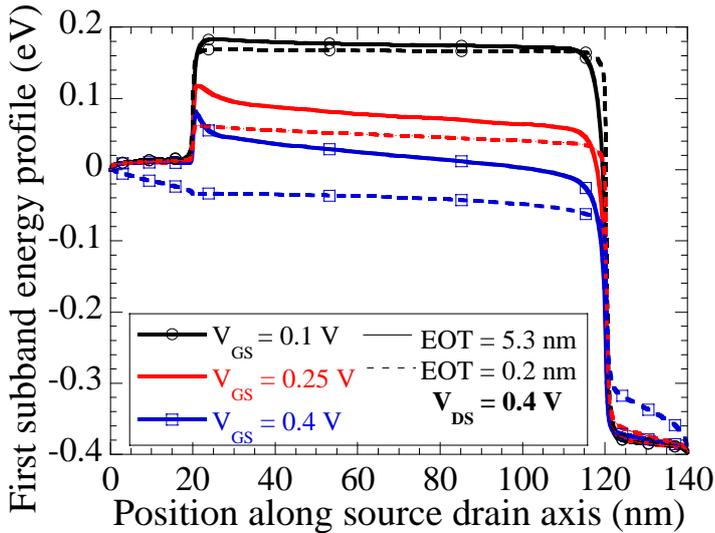
**Fig. 8.** 1st subband energy profile for two EOT values.

In Fig. 8, we can observe that the potential in the channel is all the more flat that $C_{OX}$ is higher. However, in all cases the field in the channel remains smaller than 8 kV/cm (except at the very drain-end), which should correspond to a large mean free path, higher than 100 nm (see. Fig. 5), and a high probability of ballistic transport [13].

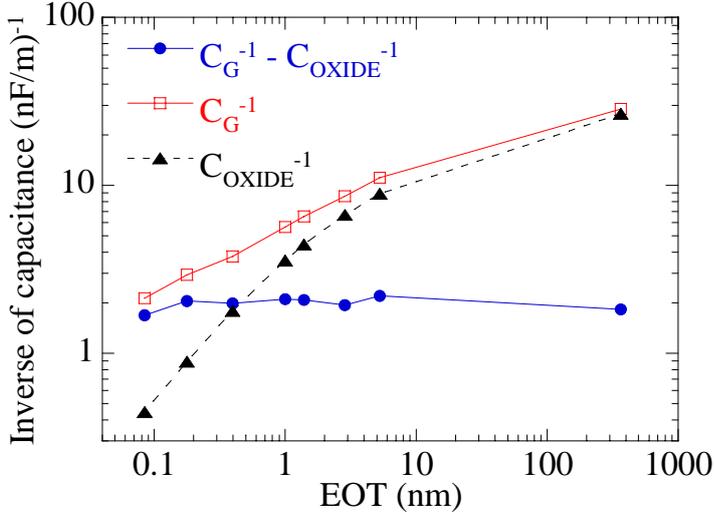

**Fig. 9.** Inverse of gate, oxide and gate minus oxide capacitances, calculated at low $V_{DS}$, as a function of equivalent oxide thickness EOT.

To understand the origin of this weak field in the channel, in Fig. 9 is plotted the evolution of the inverse of both the geometrical oxide and total gate capacitances as a function of EOT. The gate oxide capacitance is computed from the well-known expression for a cylindrical structure where the nanotube is considered as equipotential:

$$C_{OX} = 2\pi\varepsilon_{oxide} / \ln(2.t_{ox}/d_{tube} + 1) \qquad (2)$$

The total gate capacitance here corresponds to the derivative of channel charge as a function of the gate voltage $V_{GS}$ computed at low bias ($V_{DS} = 0.05$ V). This low source-to-drain voltage allows assuming a dominant gate contribution to the external electric field in the tube region. For high EOT values, the figure shows that the gate capacitance is equal to oxide capacitance. By decreasing the oxide thickness, the total gate capacitance deviates from the $C_{OX}$ trend and $1/C_G$ tends to $1/C_G - 1/C_{OX}$.

This phenomenon corresponds to the well-known concept of quantum capacitance limit regime [26,27]. It originates from the reduced density of states, associated with low dimensional structures, which cannot provide enough charge to screen the gate-induced field. The channel potential is thus fixed by the gate, especially when $C_{OX}$ is high. To account for the correct value of the induced charge, this quantum capacitance $C_Q$ must be added in the simplified expression of the total gate capacitance by the series connection of the geometric and the quantum capacitances:

$$C_G^{-1} = C_{OX}^{-1} + C_Q^{-1} \qquad (3)$$

In Fig. 9, by increasing EOT, the evolution of $1/C_Q = 1/C_G - 1/C_{OX}$ remains nearly constant and is about 4-5 pF/cm. This result agrees with the common value $C_Q = 4$ pF/cm usually considered in a (metallic) CNT with one occupied subband [28].

The weak field observed in the channel can also be related to highly ballistic transport as mentioned above. It is well illustrated in Fig. 10 where we plot the fraction of electrons as a function of the number of scattering events experienced in the channel. The result for a 100 nm-long CNTFET is compared with those of Si all-around-gated MOSFET and double gate (DG) MOSFETs of different lengths. For the latter, the distribution is a bell curve for larger length and tends to decrease exponentially by shrinking the channel length. It illustrates the cross-over between stationary transport and non-stationary or even quasi-ballistic transport. At a given channel length (100 nm), CNTFET exhibit strong ballistic characteristics compared to DG MOS Si and the fraction of ballistic electrons in the CNTFET (50%) is as high as in a 15 nm-long GAA Si FET. It should be noted that, in contrast to what is often considered in many theoretical works, electron transport is not purely ballistic: electron-phonon scattering has to be taken into account in CNTFETs operating at room temperature [11].

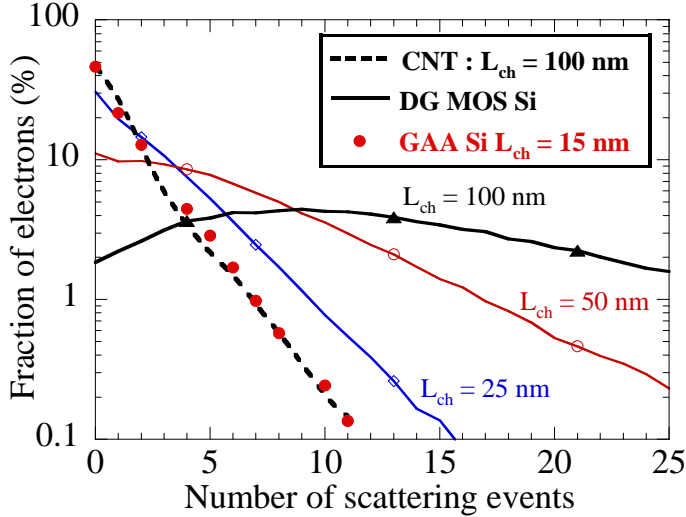

**Fig. 10.** Fraction of electrons crossing the channel as a function of the number of scattering events for $C_{OX} = 72 \text{ pF/m}$ ($V_{GS} = V_{DS} = 0.4 \text{ V}$); comparison with Si DG MOS of different lengths ($V_{GS} = V_{DS} = 0.7 \text{ V}$) and 15 nm-long Si GAA (5 nm × 5 nm cross section).

### 3. Dynamic performance for logical gate

In this part we propose to evaluate the dynamic performance for logic device applications. Some factors of merit for simulated devices with different $C_{OX}$ values are listed in Table I. We also report experimental results from Javey *et al.* obtained with a planar gate geometry and chemically-doped ohmic contacts ($C_{OX} = 280 \text{ pF/m}$ – EOT = 0.86 nm) [6]. All values are evaluated in the on-state corresponding to $V_{GS} = V_{DS} = 0.7 \text{ V}$. High values of subthreshold slope are obtained, near the theoretical limit of 60 mV/dec for lower dielectric thicknesses. The transconductances $g_m$ and the on-state currents $I_{on}$ obtained are in reasonable agreement with experimental data. However, the perfect electrostatic gate control by the coaxial geometry and the ideal character of simulated ohmic contacts can explain better performances obtained than in experimental conditions. Moreover, the simulation does not include the influence of some parasitic effects like those due to coulombic centers related to chemical doping [29].

| EOT *(nm)* | 1.4 | 0.4 | 0.2 | 0.86 (ref. [6]) |
|---|---|---|---|---|
| *S* **(mV/dec)** | 70 | 65 | 60 | 70 |
| $I_{OFF}$ **(nA)** | ≈ 0.1 | ≈ 0.1 | ≈ 0.1 | ≈ 0.3 |
| $I_{ON}$ **(μA)** | 20 | 33 | 35 | 5 |
| $g_m$ **(μS)** | 50 | 100 | 150 | 20 |

**Table I.** Device parameters extracted from simulation with different $C_{OX}$ at given $I_{OFF}$ and compared with experimental data [6].

To make easier and more relevant the comparison between CNT and Silicon-based MOSFETs, the $I_{ON}/I_{OFF}$ ratio and the intrinsic gate delay τ are convenient figures of merit, widely used to compare devices of very different geometries. We plot in Fig. 11 the $I_{ON}/I_{OFF}$ ratio as a function of the power supply voltage for three EOT values at given $I_{OFF} = 0.1$ nA. A fairly good agreement is found with the experimental result extracted from [6]. We also compare CNTFETs performance with that of two aggressively scaled Si gate-all-around (GAA) MOSFETs. Particularly, the 19 nm-long achieves a very strong $I_{ON}$ current due to a better scaling trade-off ($EOT = 1$ nm) and to optimized source-drain access resistance. The CNTFET reaches performance similar to the latter, though the length is 5 times higher, and outperforms the results obtained with the 15 nm-long device (but $EOT = 1.2$ nm), which illustrates the high potential of CNT-based transistors.

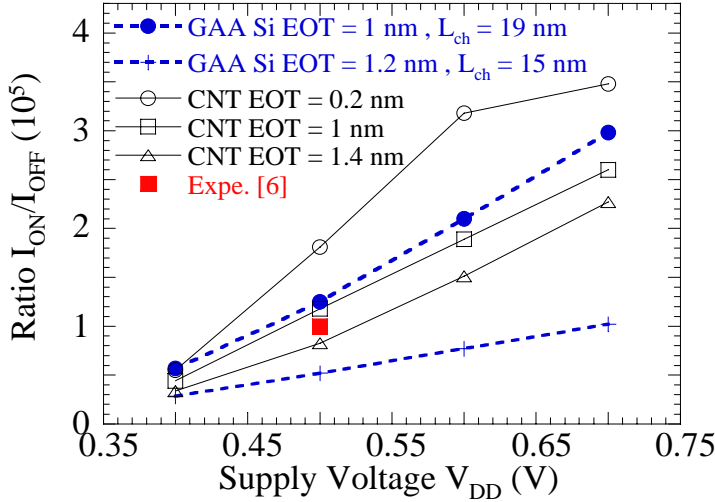

**Fig. 11.** $I_{ON}/I_{OFF}$ **ratio as a function of** $V_{DD}$ **at given** $I_{OFF} = 0.1$ **nA; comparison with experiment [6] and simulated Si 15 nm-long (5 nm by 5 nm cross section) and 19 nm-long (8 nm by 8 nm cross section) GAA-MOSFETs.**

The intrinsic gate delay characterizes how fast a transistor switches. It is a key metric for logic device applications which provides a frequency limit for transistors. It is computed as $\tau = C_G \cdot V_{DD} / I_{ON}$, where $C_G$ is the differential gate capacitance for $V_{GS} = 0.7$ V and low $V_{DS}$, and is plotted as a function of $I_{ON}/I_{OFF}$ ratio in Fig. 12. The difference of workfunction $\Delta\Phi$ between a midgap gate and another gate material is used as a parameter: the workfunction $\Phi_M$ of the gate material is equal to $\Phi_{midgap} + \Delta\Phi$. For CNTFETs, a limited range of $\Delta\Phi$ shifts is carefully considered in order to prevent from subthreshold effects like ambipolar behaviour or band-to-band tunnelling.

Surprisingly, for a CNTFET, the thicker EOT, the better the intrinsic delay. Indeed, by decreasing EOT, the combined effect of $I_{ON}$ limited enhancement and total gate capacitance saturation occurring in the quantum capacitance regime, yields an increase of τ. Now we compare CNTFET intrinsic delay with silicon-based-devices-performance deduced from Monte Carlo simulation. CNTFET performance is 10 times better than that of 100 nm-long DG Si structures. Even with a ballistic Si channel the delay of the DG-MOSFET is still higher than in CNTFETs. The latter can reach the same intrinsic delay than the 19 nm-long high-performance GAA Si-MOSFET. So by further decreasing the CNTFET channel length close to the phonon mean-free-path, we can expect to reach better intrinsic gate delay values. In [30] and [31], for 15 nm long CNTFETs, τ values of the order of 10 femtoseconds were obtained by simulation, which shows that such devices are very promising for fast logic applications.

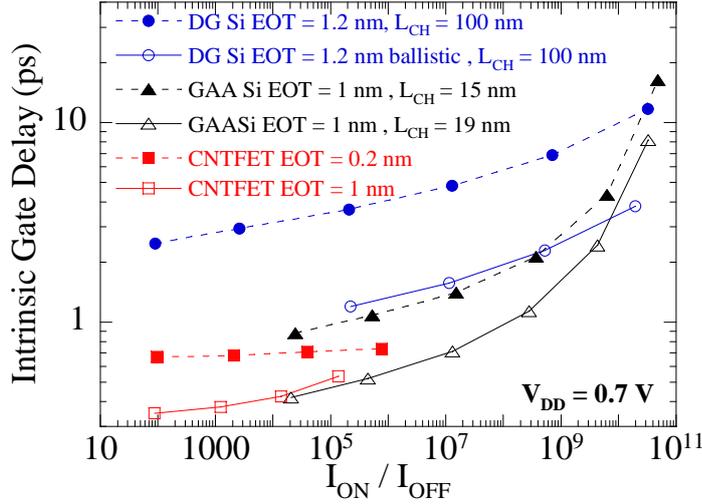

**Fig. 12.** Intrinsic gate delay as a function of $I_{ON}/I_{OFF}$ at given $V_{DD}$ = 0.7 V; comparison with simulated Si DG and GAA-MOSFET.

### 4. High-frequency performance

Now we focus on CNTFET High-Frequency (HF) performance. One of the key parameter which characterizes the high frequency potential of a transistor is the unit-gain current cut-off frequency $f_T$. We extract this cut-off frequency $f_T$ within the usual quasi static approximation where the result is given by the well known expression:

$$f_T = \left.\frac{g_m}{2\pi C_G}\right|_{V_{DS}=V_{GS}=V_{DD}} \qquad (4)$$

where $g_m$ is the transconductance and $C_G$ is the intrinsic gate capacitance. They are obtained by derivating the drain current and the channel charge with respect to the gate voltage, for $V_{DS} = V_{DD}$, respectively. According to the small signal model of a CNT [7,32], we carefully consider the drain current range between threshold and saturation for computing $g_m$, to avoid source-drain access resistance influence.

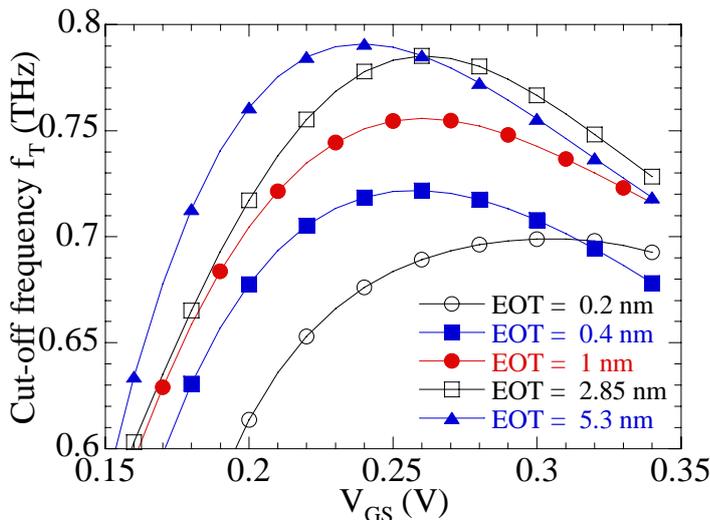

**Fig. 13.** Cutoff frequency as a function of $V_{GS}$ for different EOT values for $V_{DS}$ = 0.7 V.

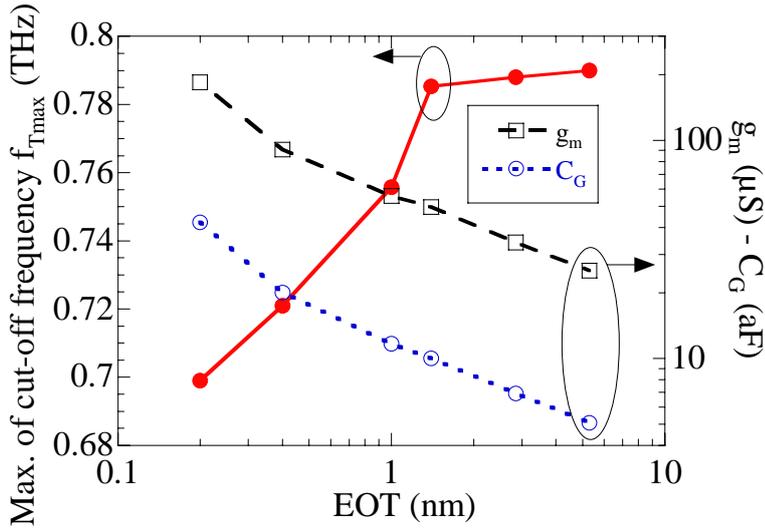

**Fig. 14.** Maximum of the cut-off frequency $f_{Tmax}$, $g_m$ and $C_G$ as a function of EOT.

Fig. 13 plots the CNTFET cutoff frequency as a function of $V_{GS}$ for different EOT values. For each EOT, $f_T$ follows a bell shaped curve. By considering the higher $f_T$ values which ranges between 0.2 V and 0.3 V, we can extract maximum cut-off frequency value $f_{Tmax}$ that we plot in Fig. 14 as a function of EOT. This curve can be divided into two parts. For low EOT values (EOT< 1.4 nm), when rising EOT we can observe a slight improvement of $f_{Tmax}$ which tends to saturate for higher EOT values. This particular behaviour can be understood with the help of $g_m$ (dashed line) and $C_G$ (dotted line) evolutions, both plotted in Fig. 14. Due to the influence of the quantum capacitance at low EOT values, $g_m$ decreases more slowly than $C_G$ when increasing EOT which yields the enhancement of $f_T$ observed in Fig. 14. For higher EOT, the quantum capacitance effects are negligible, so that $g_m$ and $C_G$ decrease in parallel and $f_{Tmax}$ becomes EOT-independent.

| Burke [32] | Hasan [33] | Guo [30] | This work | Le Louarn [7] |
|---|---|---|---|---|
| Ballistic | Ballistic | Scattering | Scattering | Expe. |
| EOT= X | EOT= 0.4 | EOT= 1.2 | EOT= 1.2 | EOT≈1.3 |
| **800GHz** | **1.3THz** | **400GHz** | **780GHz** | **30GHz** |

**Table II.** Device parameters extracted from simulation with different $C_{OX}$ for $L_{ch}$ = 100 nm and compared with experimental data [7].

In Table II are reported some of the most significant simulated and experimental state-of-the-art results compared to ours. Burke [32] and Hasan [33] proposed an equivalent circuit in the ballistic limit of operation and estimated $f_T$ to be 80 GHz and 130 GHz divided by the tube length in micrometers, respectively. Guo [30] examined the effect of phonon scattering and channel length on the unit-current-gain frequency $f_T$ and fitted by $f_T = 110\text{ GHz.µm}/L_{ch}$ and $f_T = 40\text{ GHz.µm}/L_{ch}$ in ballistic and scattering cases, respectively. Our results are in agreement with ballistic results. The lower performance obtained by Guo [30] is probably due to the effect of Schottky contacts. Only few experimental studies were able to assess high frequency performance and small-signal voltage gain, due to the low output current of a single-carbon nanotube and to parasitic capacitances associated with wide contacts. These problems were overcome in [7] by increasing the density of nanotubes and the number of gate fingers. However, in spite of high transconductance values ($g_m = 500\text{ µS}$), an intrinsic $f_T$ value of 30 GHz was extracted from extrinsic measurements by deembedding parasitic capacitances influence. It constitutes the best $f_T$ value ever measured but remains far away from theoretical and simulated predictions for ideal CNTFET devices. Combining CNTFET-fabrication improvement and simulation may give key-parameters to improve the dynamical behaviour and to bring $f_T$ towards a higher level of performance.

## 5. Conclusion

The particle Monte Carlo technique including an accurate description of electron-phonon scattering was used to analyze the transport properties of semiconducting CNTs and the static and dynamic performance of 100 nm-gate length CNTFET. Besides the high intrinsic mobility according to their diameter, CNTs are shown to have a high potential of ballistic transport on large distance in low-field regime limited by intravalley acoustic phonon scattering with mean-free path of more than 100 nm. Beyond a field of 10 kV/cm the transport is dominated by high energy intervalley phonons, which reduce the mean-free path up to 20 nm for a field of 100 kV/cm.
The 3D self-consistent study of CNT-based transistors with ohmic source and drain contacts shows the high level of performance that can be expected from such devices that may operate close to the quantum capacitance regime for equivalent gate oxide thicknesses lower than 1 nm. Combining the excellent electrostatic gate control with the high fraction of ballistic electrons, very good $I_{ON}/I_{OFF}$ ratio and intrinsic delay $\tau = C_G V_{DD}/I_{ON}$ may be reached for 100 nm-long devices with performance comparable to that of Silicon-based transistors of gate length smaller than 20 nm. A maximum intrinsic cut-off frequency $f_T$ of 780 GHz is predicted for an EOT of 1.2 nm, which strongly suggests that $f_T$ of 1 THz could be reached for smaller gate length.

## Acknowledgment

This work was supported by the European Community through Network of Excellence SINANO and the French ANR through PNANO HFCNT project.

# References

[1] A. Javey et al., High-Field Quasiballistic Transport in Short Carbon Nanotubes, Phys. Rev. Lett. 92 (2004) 106804.
[2] J. Chen et al., Self-aligned carbon nanotube transistors with novel chemical doping, in Proc. IEEE IEDM Tech. Dig. 695 (2004).
[3] T. Durkop et al., Extraordinary mobility in semi-conducting carbon nanotubes, Nano Lett. 4-35 (2004).
[4] M. Radosavljevic et al., High performance of potassium n-doped carbon nanotube field-effect transistor , Appl. Phys. Lett. 84 (2004) 3693.
[5] Y. M Lin et al., High-performance dual-gate carbon nanotube FETs with 40-nm gate length, IEEE Electron Dev. Lett. 26 (2005) 823.
[6] A. Javey et al., High performance n-type carbon nanotube field-effect transistors with chemically doped contacts, Nano Lett. 5 (2005) 345.
[7] A. Le Louarn et al., Intrinsic current gain cutoff frequency of 30 GHz with carbon nanotube transistors, Appl. Phys. Lett. 90 (2007) 233108.
[8] B. M. Kim et al., High-performance carbon nanotube transistors on SrTiO/Si substrates, Appl. Phys. Lett. 84 (2004) 1946.
[9] J. Appenzeller et al., Band-to-Band Tunneling in Carbon Nanotube Field-Effect Transistors, Phys. Rev. Lett. 93 (2004) 196805.
[10] A. Javey et al., Ballistic Carbon Nanotube Field-Effect Transistors, Nature 424 (2003) 654.
[11] J. Guo and M. Lundstrom, Role of phonon scattering in carbon nanotube field-effect transistors, Appl. Phys. Lett. 86 (2005) 193103.
[12] P. Dollfus et al., Effect of discrete impurities on electron transport in ultra-short MOSFET using 3D Monte Carlo simulation, IEEE Trans. Electron Dev. 51 (2004) 749.
[13] H. Cazin et al. Electron-phonon scattering and ballistic behavior in semiconducting carbon nanotubes, Appl. Phys. Lett. 87 (2005) 172112.
[14] R. Saito et al., in Physical properties of carbon nanotubes (Imperial College Press, London,), 1988.
[15] T. Hertel and G. Moos, Electron-Phonon Interaction in Single-Wall Carbon Nanotubes: A Time-Domain Study, Phys. Rev. Lett. 84 (2000) 5002.
[16] G. Pennington and N. Goldsman, Semi-classical transport and phonon scattering of electrons in semiconducting carbon nanotubes, Phys. Rev. B 68 (2003) 045426.
[17] O. Dubay, G. Kresse, Accurate density functional calculations for the phonon dispersion relations of graphite layer and carbon nanotubes, Phys. Rev. B 67 (2003) 035401.
[18] A. Verma et al., Effects of radial breathing mode phonons on charge transport in semiconducting zigzag carbon nanotubes, Appl. Phys. Lett. 87 (2005) 123101.
[19] M. Machon et al., Strength of radial breathing mode in single-walled carbon nanotubes, Phys. Rev. B 71 (2005) 035416.
[20] V. Perebeinos et al., Electron-Phonon Interaction and Transport in Semiconducting Carbon Nanotubes, Phys. Rev. Lett. 94 (2005) 086802.
[21] J. Saint Martin, A. Bournel, and P. Dollfus, On the ballistic transport in nanometer-scaled DG MOSFETs, IEEE Trans. Electron Dev. 51 (2004) p. 1148.
[22] P. Lugli and D. K. Ferry, Degeneracy in the Ensemble Monte Carlo Method for High-Field Transport in Semiconductors, IEEE Trans. Electron Devices 32 (1985) 2431.
[23] Y.-M.Lin, et al., High-performance carbon nanotube field-effect transistor with tunable polarities, IEEE Trans. Nanotechnol. 4 (2005) 481.
[24] V. Derycke, et al., Controlling doping and carrier injection in carbon nanotube transistors, Appl. Phys. Lett. 80 (2002) 2773.
[25] J. Chen et al., Self-aligned carbon nanotube transistors with charge transfer doping, Appl. Phys. Lett. 86 (2005) 123108.
[26] S. Datta, Electronic transport in mesoscopic systems, Cambridge University Press, (1995).
[27] S. Luryi, Quantum capacitance devices, Appl. Phys. Lett. 52 (1988) 501.
[28] D.L. John et al., Quantum capacitance in nano-scale device modelling, J. Appl. Phys. 96 (2004) 5180.

[29] A. G. Petrov et al., Transport in nanotubes: Effect of remote impurity scattering, Phys. Rev. B 70 (2004) 035408.
[30] J. Guo et al., Effect of Phonon Scattering on Intrinsic Delay and Cutoff Frequency of Carbon Nanotube FETs, IEEE Transactions on Electron Devices 53-10 (2006) 2467-2469.
[31] G. Fiori and G. Iannaccone, A three-dimensional simulation study of the performance of carbon nanotube field-effect transistors with doped reservoirs and realistic geometry, IEEE Transactions on electron devices 53-8 (2006) 1782-1788.
[32] P. J. Burke, AC performance of nanoelectronics: towards a ballistic THz nanotube transistor", Solid State Electron 48-11 (2004) 1981-1986
[33] S. Hasan High-frequency performance projections for ballistic carbon-nanotube transistors, IEEE Transaction on nanotechnology vol.50, no. 1 (2006) 14-22.